\documentclass[11pt]{article}
\usepackage{moriond,epsfig}

\usepackage{amssymb}

\def\be{\begin{equation}}
\def\ee{\end{equation}}
\def\bea{\begin{eqnarray}}
\def\eea{\end{eqnarray}}

\def\Lambdaqcd{\Lambda_{\ensuremath{\it QCD}}}
\newcommand{\sqrts}{\sqrt{s}}
\newcommand{\sqrtsnn}{\sqrt{s_{_{NN}}}}
\newcommand{\sqrtsgN}{\sqrt{s_{_{\gamma N}}}}
\newcommand{\sqrtseN}{\sqrt{s_{_{e N}}}}

\providecommand{\ups}{\Upsilon}

\providecommand{\dNdeta}{dN_{ch}/d\eta|_{\eta=0}}

\def\ttt#1{\texttt{\small #1}}

\providecommand{\gpb}{\gamma\,$Pb$}

\begin{document}
\title{Experimental tests of small-$x$ QCD}

\author{ David d'Enterria}
\address{CERN, PH-EP \\
CH-1211 Geneva 23, Switzerland}

\maketitle
\abstracts{
Current and future experimental studies of the high-energy limit of QCD, dominated 
by non-linear gluon dynamics in the low-$x$ sector of the hadron wavefunctions, are presented. 
Results at HERA (proton) and RHIC (nucleus) pointing to the possible onset of parton 
saturation phenomena, and perspectives at the LHC and new proposed DIS 
facilities are outlined.}
\noindent
{\small¥{\it Keywords}: QCD, low-$x$, gluon saturation, HERA, RHIC, LHC.}


\section{Introduction}

\begin{figure}[htb]
\begin{center}
\hspace*{-.5cm}
\epsfig{file=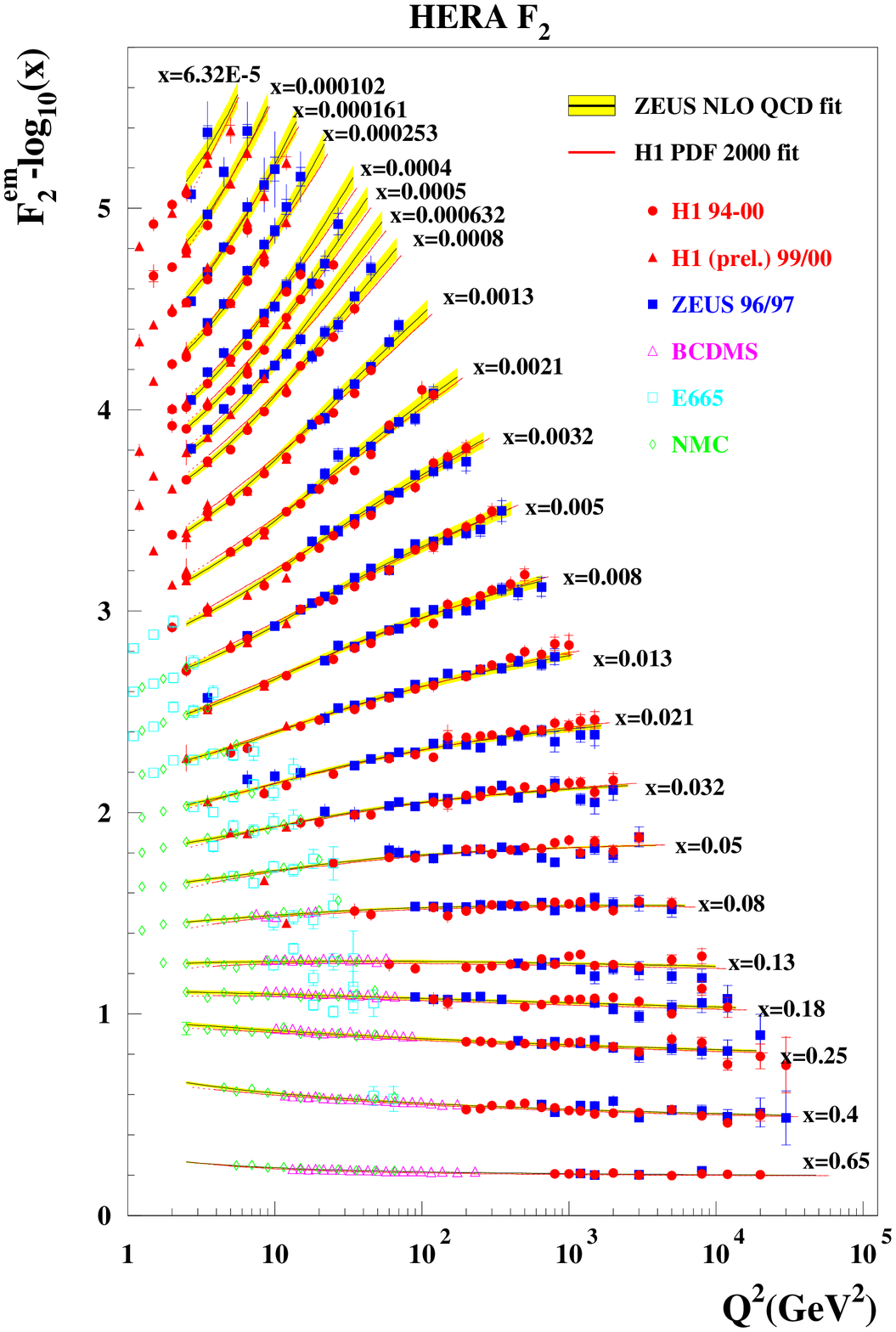,width=8.5cm,height=8.0cm}
\epsfig{file=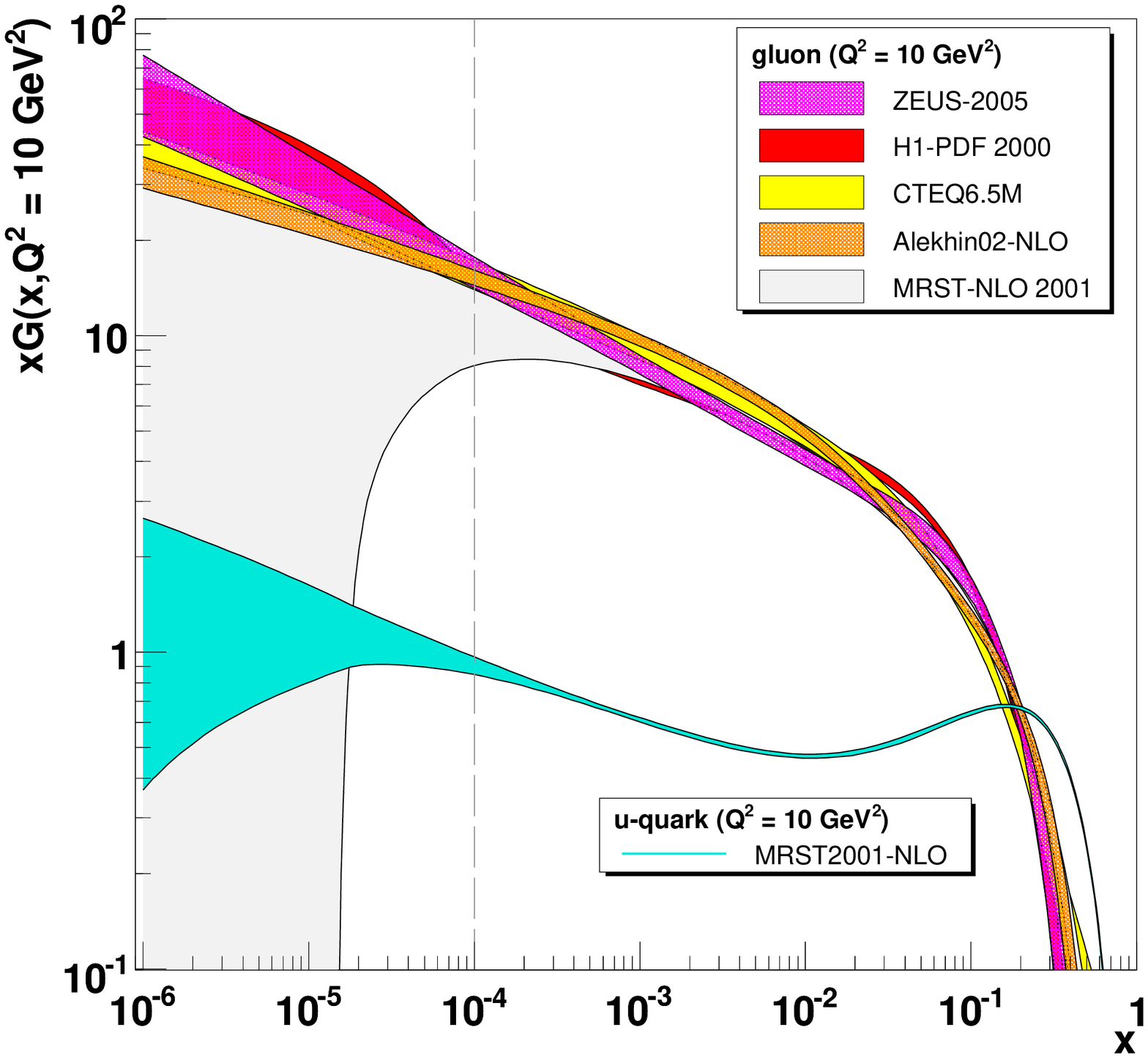, width=8.cm,height=7.7cm}
\caption{Left: Structure function $F_2(x,Q^2)$ measured in proton DIS at HERA ($\sqrts$ = 320 GeV) 
and fixed-target ($\sqrts\approx$ 10-30 GeV) experiments. Right: Gluon distribution function in the
proton, $xG(x,Q^2$ = 5 GeV$^2)$, derived from DGLAP-based analyses of $F_2$ scaling violations 
(for comparison, the bottom curve shows the $u$ quark PDF).}
\label{fig:HERA_F2_xG}
\end{center}
\end{figure}

Gluons provide the dominant contribution to hadronic scattering cross sections 
at high-energies. Deep-Inelastic (DIS) experiments of electrons off protons at the 
HERA collider at DESY have shown that for values of the parton momentum fraction 
$x=p_{\mbox{\tiny{\it parton}}}/p_{\mbox{\tiny{\it proton}}} \lesssim 0.01$, the 
proton wavefunction is basically purely gluonic (Fig.~\ref{fig:HERA_F2_xG}). This is so, 
because gluons are ``cheap'' to radiate: the probability of emitting a gluon increases as 
$\propto\alpha_s\ln(Q^2)$ and $\propto\alpha_s\ln(1/x)$ according to the standard 
linear QCD evolution (DGLAP~\cite{dglap} and BFKL~\cite{bfkl} resp.) equations. 
As a matter of fact, for decreasing values of $x$ the gluon density increases so fast
that unitarity would be ultimately  violated, even for processes with large virtualities 
$Q^2\gg \Lambdaqcd^2$. The theoretical expectation~\cite{GLR,MQ} is that at some 
small enough value of $x$ ($\alpha_s\ln(1/x) \gg 1$) 
non-linear gluon-gluon fusion effects -- not accounted for in the DGLAP/BFKL equations 
-- will become important and will tame the growth of the parton densities. 
The onset of saturation in the proton (or in a nucleus with $A$ nucleons) is expected
for parton momenta $Q^2\lesssim Q^2_s$ where $Q_s$ is a dynamical ``saturation 
scale''~\cite{GLR,MQ} which depends on the transverse size ($\pi R^2$) of the hadron:
\begin{equation}
Q_s^2(x)\simeq \alpha_s \frac{1}{\pi R^2}\,xG(x,Q^2)\sim A^{1/3}\,x^{-\lambda} \sim A^{1/3}(\sqrts)^{\lambda} \sim A^{1/3}e^{\lambda y},
\label{eq:Qs}
\end{equation}
with $\lambda\approx$ 0.25~\cite{kharzeev_kln}. Eq.~(\ref{eq:Qs}) tell us that $Q_s$ 
grows with 
the energy of the collision, 
$\sqrts$, or equivalently, with the rapidity of the parton $y=\ln(1/x)$. The nucleon number 
$A$ dependence implies that, at equivalent energies, saturation effects will be amplified by 
factors as large as $A^{1/3}\approx$ 6 in heavy nuclear targets ($A$ = 208 for Pb) 
compared to protons. The regime of high gluon densities is often described in terms of the 
colour-dipole~\cite{mueller90,gbw} or ``Colour Glass Condensate'' (CGC)~\cite{cgc} 
effective theories, with the corresponding non-linear BK/JIMWLK~\cite{bk,jimwlk} evolution 
equations. 

\section{Gluon saturation at HERA ?}

Although the arguments for saturation are well justified theoretically, no strong deviation from the 
linear QCD equations has been conclusively observed in the perturbative kinematical range covered at HERA. 
Most of the experimental observables, in particular those of more inclusive nature such as the total 
$\sigma_{\gamma^*p}$ cross-section $d^2\sigma/dx\,dQ^2 \approx 2\pi\alpha^2/(x\,Q^4)F_2(x,Q^2)$,
are in good accord with the standard DGLAP expectations (Fig.~\ref{fig:HERA_F2_xG}). However, it is 
worth to note that 
the saturation scale at HERA energies is in a regime of relatively low virtualities, $Q_s^2\approx$ 1 GeV$^2$
and thus -- since Bjorken $x$ and virtuality are correlated as $x \approx Q^2/s$ -- the interpretation of most of the 
truly low-$x$ range probed ($x\lesssim 10^{-4}$) is ``blurred'' by its proximity to the non-perturbative regime.
As a consequence, the gluon distribution function $xg(x,Q^2)$ indirectly obtained from the $F_2$ scaling violations 
-- via $\partial F_2(x,Q^2)/\partial \ln (Q^2) \approx 10\,\alpha_s(Q^2)/(27\pi)\,xg(x,Q^2)$ -- 
is poorly constrained below $x\approx$ 10$^{-4}$.
Different DGLAP parametrizations\footnote{See \ttt{http://durpdg.dur.ac.uk/hepdata/pdf3.html}} 
based on DIS-only data (ZEUS-PDF, H1-PDF, Alekhin02) or on global-fits (CTEQ6.5M, MRST-NLO) 
yield gluon PDFs differing by factors of 3 or more (Fig.~\ref{fig:HERA_F2_xG}, right).\\

\begin{figure}[htb]
\begin{center}
\hspace*{.5cm}
\centerline{
\epsfig{file=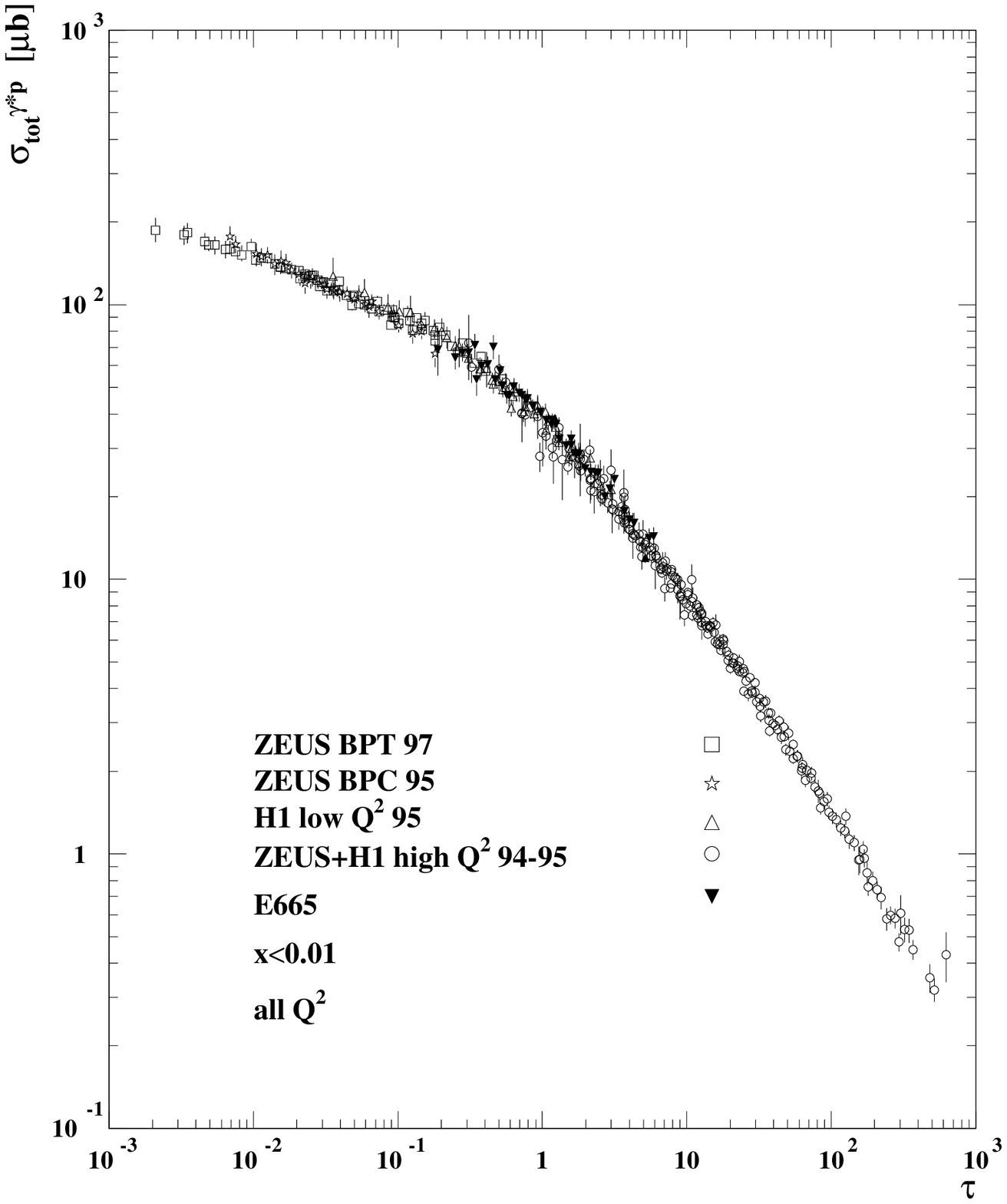,bb=0 0 450 560,clip=true,width=5.8cm,height=6.2cm}
\epsfig{file=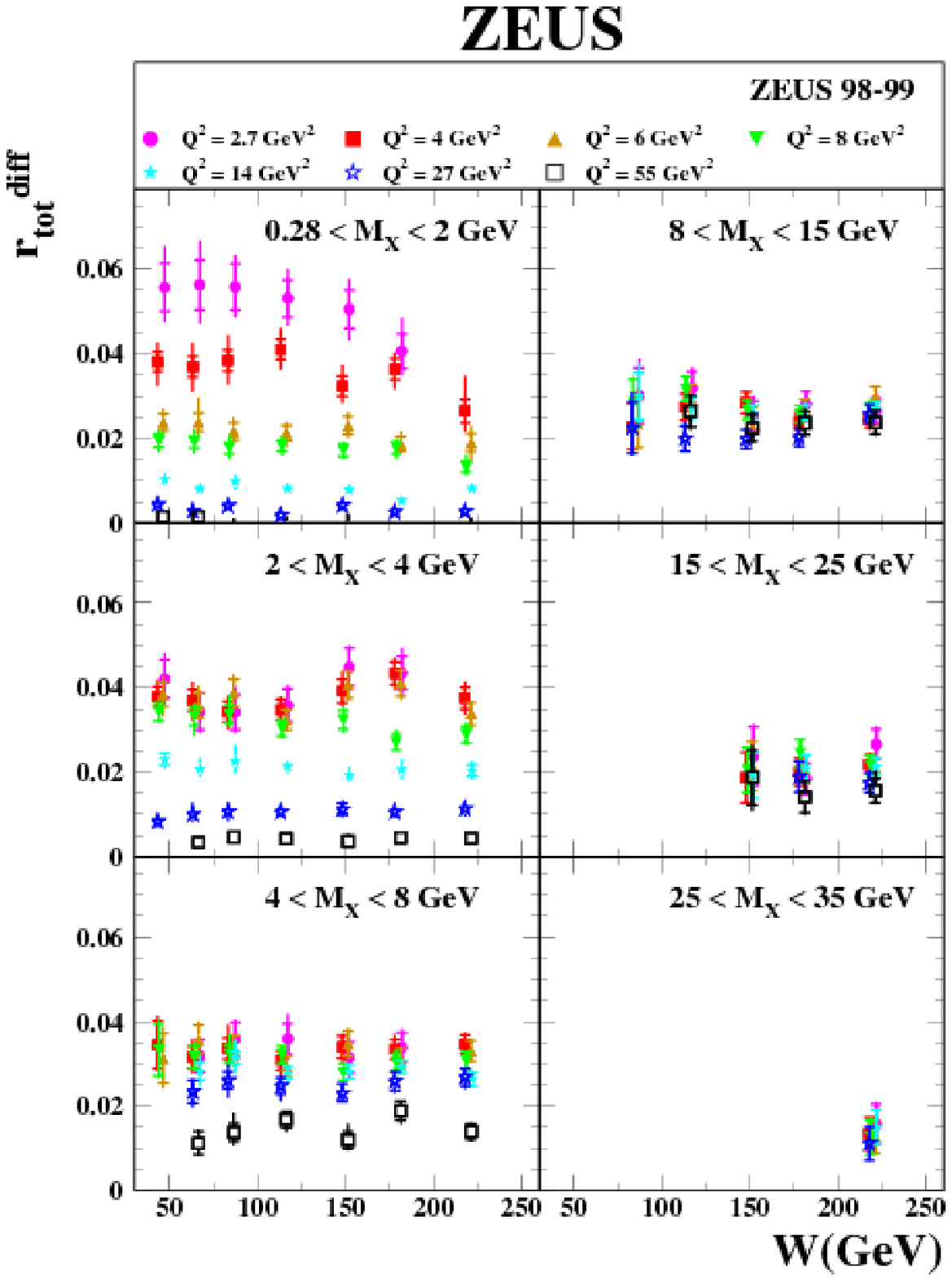,width=6.2cm,height=6.3cm}
\epsfig{file=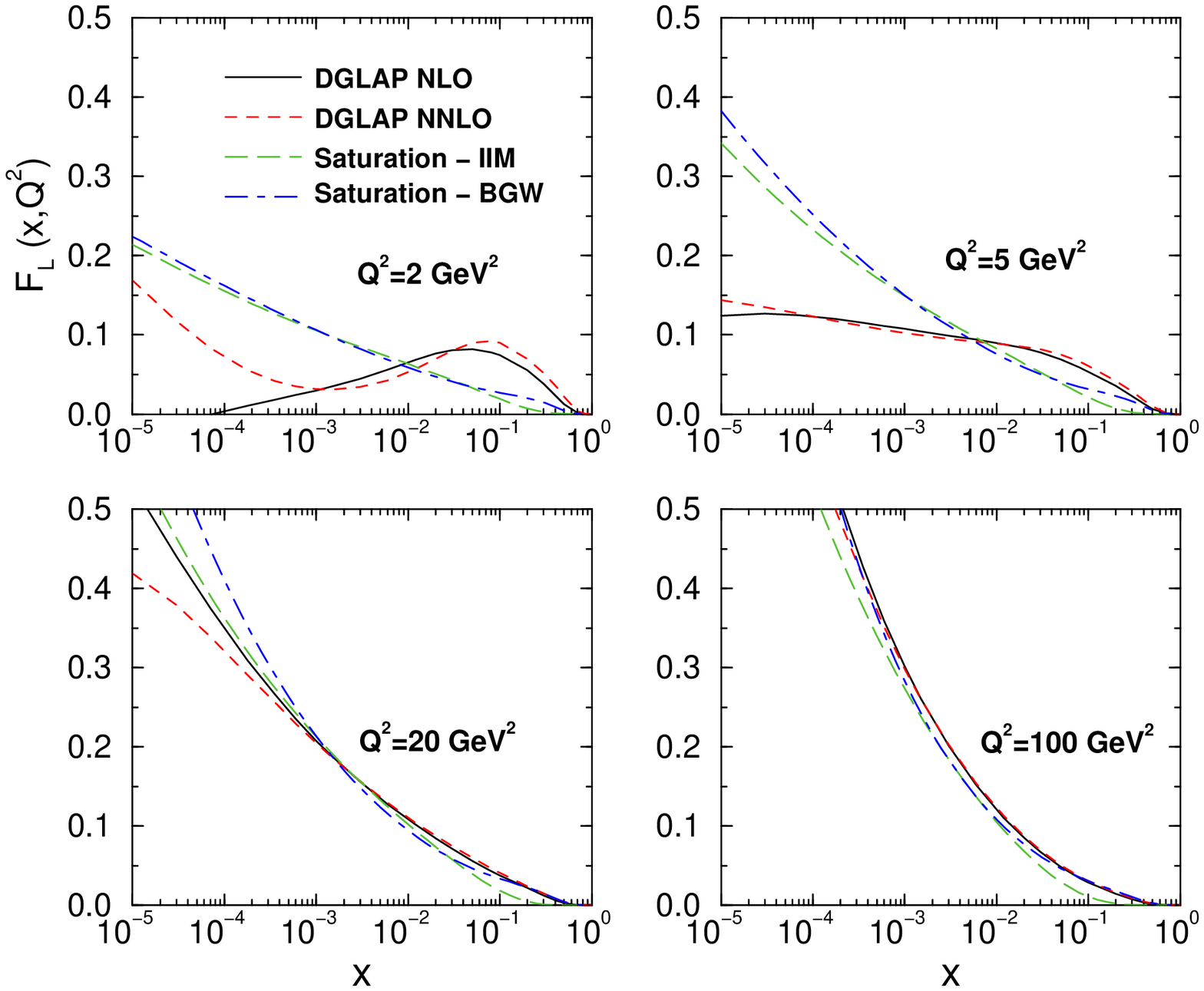,bb=19 260 293 469,clip=true,width=6.cm,height=6.cm}
}
\caption{Hints of saturation at HERA. Left: Geometric scaling~\protect\cite{gbw} of $\sigma_{\gamma*p}$ 
as a function of $\tau=Q^2/Q_s^2$ for $x < 0.01$. 
Center: Ratio of diffractive to total DIS cross-sections~\protect\cite{zeus_diff_05} as a function of $W$ and $Q^2$. 
Right: Longitudinal structure function, $F_L(x,Q^2$= 2 GeV$^2)$, predicted by DGLAP~\protect\cite{mst_FL_2006}
and saturation~\protect\cite{gbw,iim,sat_FL_2005} models.}
\label{fig:hera_sat}
\end{center}
\end{figure}

\noindent
Notwithstanding those uncertainties, there are three empirical observations at HERA (summarized in 
Fig.~\ref{fig:hera_sat}) that favour a possible onset of low-$x$ parton saturation in the proton. 
The leftmost plot shows the \underline{geometric scaling}~\cite{gbw} property of inclusive $\sigma_{\rm DIS}$ 
which, instead of being a function of $x$ and $Q^2$ separately, {\bf for $\mathbf{x<0.01}$} it features a single 
dependence on the parameter  $\tau\!=\!Q^2/Q_s^2(x)$ where 
$Q_s(x)=Q_0(x/x_0)^{\lambda}$ with $\lambda\sim$ 0.3, $Q_0$ = 1 GeV, and 
$x_0\sim$ 3$\cdot$10$^{-4}$. 
Such a scaling property 
is naturally explained by gluon saturation models, whereas the DGLAP approach can only
reproduce it via a fine tuning of the initial parameterization of the gluon distribution used.
Another piece of evidence for saturation effects at HERA is provided by diffractive 
processes, where the proton remains intact after the ``quasi-elastic'' interaction with
the photon. Diffractive scattering, accounting for 10--15\% of the total DIS cross-section, 
is characterized by colourless two-gluon exchange, 
and thus it constitutes a sensitive probe of the gluon densities~\cite{arneodo_diffract}. 
Surprisingly, the \underline{ratio of the diffractive to total} $\sigma_{\gamma^*p}$ cross-sections
is found~\cite{zeus_diff_05}  to be roughly constant as a function of the center-of-mass energy $W$ 
and $Q^2$ (Fig.~\ref{fig:hera_sat}, center). This is in disagreement with the naive pQCD 
expectations of an {\it increase} of the ratio according to 
$r^{\rm diff}_{\rm tot} = \sigma_{\rm diff}/\sigma_{\rm tot}\propto |xg(x,Q^2)|^2/xg(x,Q^2)\sim 
W^{4\lambda}/W^{2\lambda}\sim W^{2\lambda}$. 
The last indication of a ``tension'' between the standard linear QCD equations and low-$x$
HERA data comes from the \underline{longitudinal structure function}, $F_L(x,Q^2)$ which
at variance with $F_2$, is directly proportional to $xg(x,Q^2)$. The $F_L(x,Q^2)$ derived 
from NLO DGLAP analyses becomes unphysically negative~\cite{mst_FL_2006} below 
$x\approx$ 10$^{-4}$ for relatively small $Q^2$ values, whereas it is a well-behaved
object in saturation models (Fig.~\ref{fig:hera_sat}, right).

\section{Gluon saturation at RHIC ?}

\begin{figure}[htb]
\begin{center}
\epsfig{file=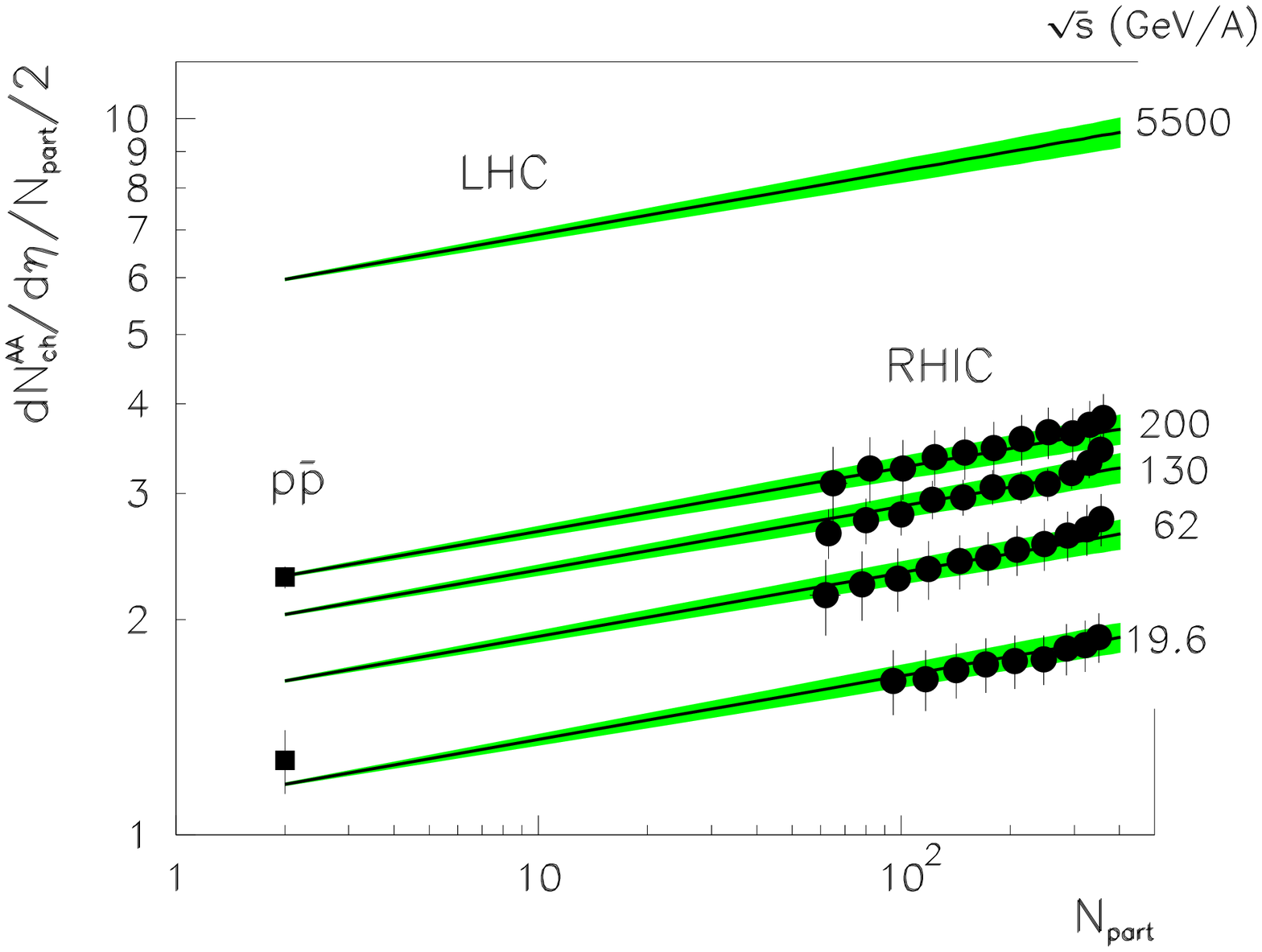,width=7.7cm}
\epsfig{file=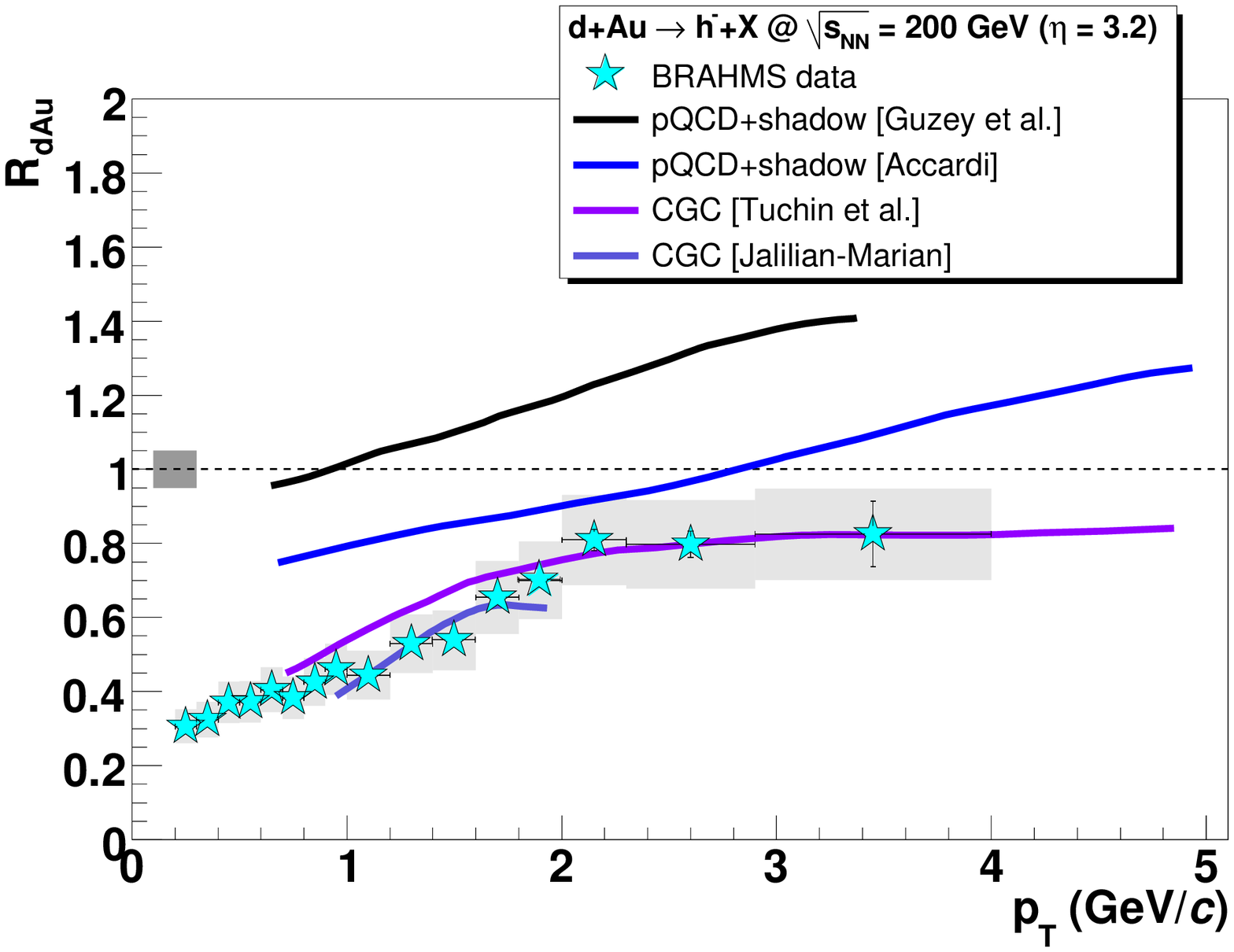,width=7.7cm}
\caption{Hints of saturation at RHIC. Left: Normalized $\dNdeta$ as a function of c.m. energy 
and centrality (given in terms of the number of nucleons participating in the collision, $N_{\rm part}$) 
measured by PHOBOS in Au-Au~\protect\cite{phobos_wp} compared with saturation predictions~\protect\cite{armesto04}.
Right: Nuclear modification factor $R_{dAu}(p_T)$ for negative hadrons at $\eta$~=~3.2 in d-Au at $\sqrtsnn$ = 200 GeV: 
BRAHMS data~\protect\cite{brahms_wp} compared to pQCD~\protect\cite{guzey04,accardi04} and 
CGC~\protect\cite{tuchin04,jamal04} predictions.}
\label{fig:rhic_sat}
\end{center}
\end{figure}

Among the interesting observations in nucleus-nucleus (A-A) collisions at RHIC is the possible onset 
of parton saturation phenomena. Though nuclei at RHIC are probed at lower energies ($\sqrtsnn$ = 200 GeV) 
than protons at HERA, saturation effects are ``amplified'' thanks to the increased transverse parton density 
in the former compared to the latter. Two empirical observations support the Color-Glass-Condensate (CGC)
predictions of a reduced parton flux 
in the incoming ions due to enhanced non-linear QCD effects. On the one hand, the measured 
hadron multiplicities~\cite{phobos_wp,phenix_wp,star_wp,brahms_wp},
$\dNdeta\approx$ 700, are significantly lower than the $\dNdeta\approx$ 1000 values predicted by 
minijet~\cite{hijing} or Regge~\cite{dpm} models, but are well reproduced by CGC approaches~\cite{mclv}. 
Parton multiplicity distributions at high energies are perturbatively calculable in saturation approaches,
since they are governed by a semihard saturation scale $Q_s^2 \propto \sqrtsnn^{\;\lambda}$ 
with an exponent $\lambda$ constrained by e-A data~\cite{armesto04}. Assuming parton-hadron
duality, hadron multiplicities at mid-rapidity rise proportionally to $Q_s^2$ times the transverse 
(overlap) area, a feature that accounts naturally for the experimentally observed factorization of 
$\sqrtsnn$- and centrality-dependences in $\dNdeta$ (Fig.~\ref{fig:rhic_sat}, left).\\

\noindent 
The second manifestation of CGC-like effects in the RHIC data is the BRAHMS observation~\cite{brahms_wp} 
of suppressed yields of semi-hard hadrons ($p_T\approx 2 - 4 $ GeV/$c$) in d-Au relative to p-p 
collisions at increasingly forward rapidities (up to $\eta\approx$ 3.2, Fig.~\ref{fig:rhic_sat}, right). 
Hadron production at such small angles is theoretically sensitive to partons in the Au nucleus 
with  $x_2^{min} = (p_T/\sqrtsnn)\,\exp(-\eta)\approx\mathcal{O}$(10$^{-3}$)~\cite{guzey04}.
The observed nuclear modification factor, $R_{dAu}\approx$ 0.8, cannot be reproduced by pQCD
calculations~\cite{guzey04,accardi04,deflorian03} that include the same {\it leading-twist} nuclear shadowing
that describes the d-Au data at $y=$ 0, but can be described by CGC approaches that parametrise 
the Au nucleus as a saturated gluon wavefunction~\cite{tuchin04,jamal04}. 


\section{Low-$x$ QCD studies at the LHC}

\noindent
The Large Hadron Collider (LHC) at CERN will provide p-p, p-A and A-A collisions 
at $\sqrtsnn$ = 14, 8.8 and 5.5 TeV  with luminosities $\mathcal{L}\sim$  10$^{34}$, 
10$^{29}$ and 5$\cdot$10$^{26}$ cm$^{-2}$ s$^{-1}$ respectively. Following Eq.~(\ref{eq:Qs}), 
the relevance of low-$x$ QCD effects will be significantly enhanced due to the increased: center-of-mass 
energy, nuclear radius ($A^{1/3}$), and rapidity of the produced partons~\cite{yr_lhc_pdfs,dde_lowx}. 
At the LHC, the saturation momentum $Q_s^2\approx$ 1 GeV$^2$ (proton) -- 5 GeV$^2$ (Pb) will be 
more clearly in the perturbative regime~\cite{kharzeev_kln}, hard probes will be copiously produced, and the $x$ 
values experimentally accessible will be much lower than at previous colliders:  $x_{2}\approx 10^{-3}(10^{-6})$ 
at central (very forward) rapidities. 
All LHC experiments have interesting detection capabilities in the forward direction which will help to constrain 
the PDFs in the very low-$x$ regime: (i) CMS~\cite{loi_cms_totem,dde_qm06} can measure inclusive jet and Drell-Yan 
production down to $x\sim 10^{-6}$ using the CASTOR calorimeter at rapidities $5.5<\eta<6.6$
as well as Mueller-Navelet dijets (very sensitive to non-DGLAP evolution)~\cite{mueller_navelet,bfkl_lhc} 
separated by rapidities as large as $\Delta\eta\sim$ 10,
(ii) ALICE and LHCb feature a forward muon spectrometer (covering $2\lesssim\eta\lesssim 5$)
which gives them access to heavy-quark, quarkonia and gauge boson measurements~\cite{Baines:2006uw} 
down to $x\sim 10^{-5}$.\\

\begin{figure}[htbp]
\epsfig{figure=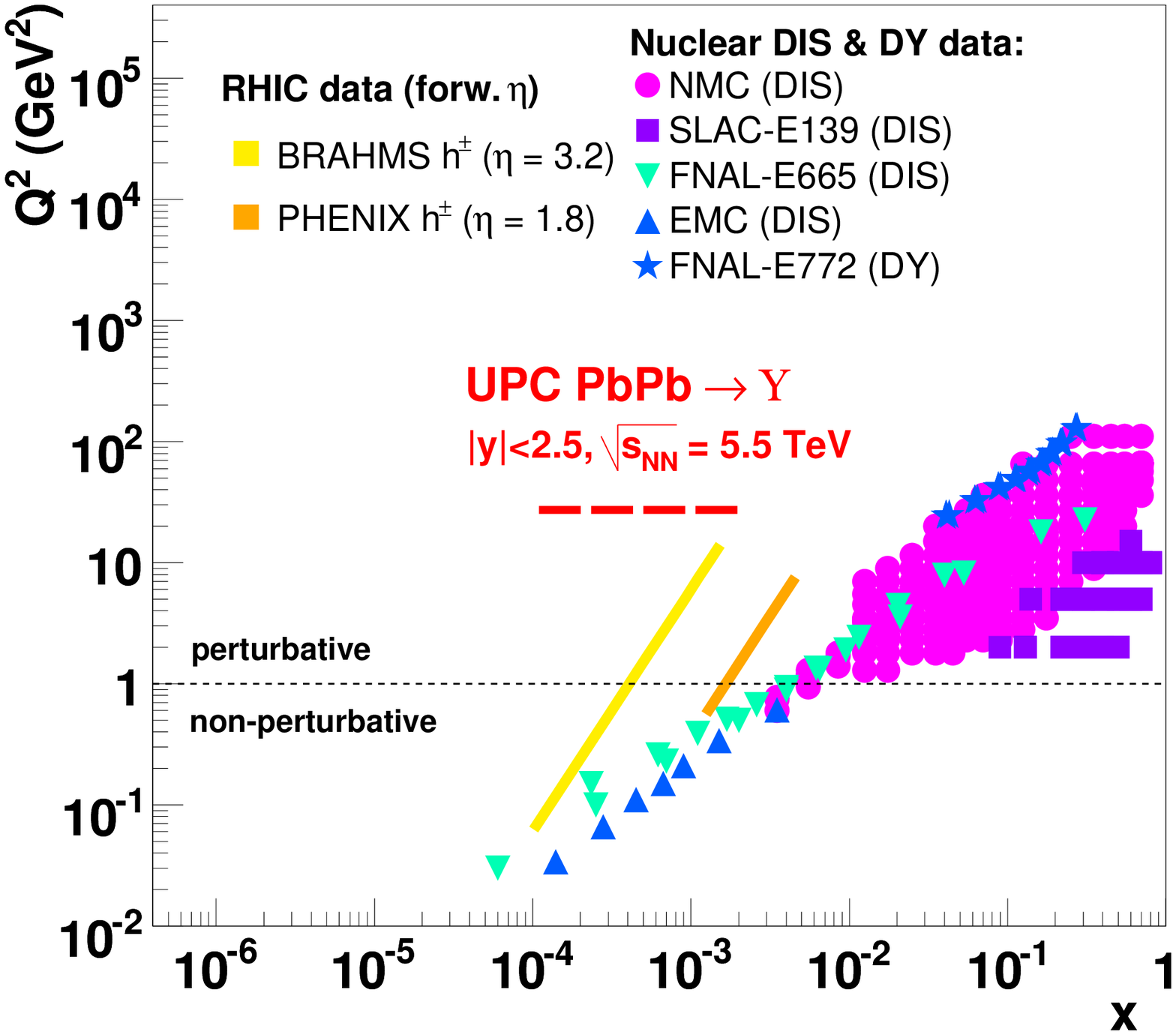,height=7.9cm,width=7.9cm}
\epsfig{figure=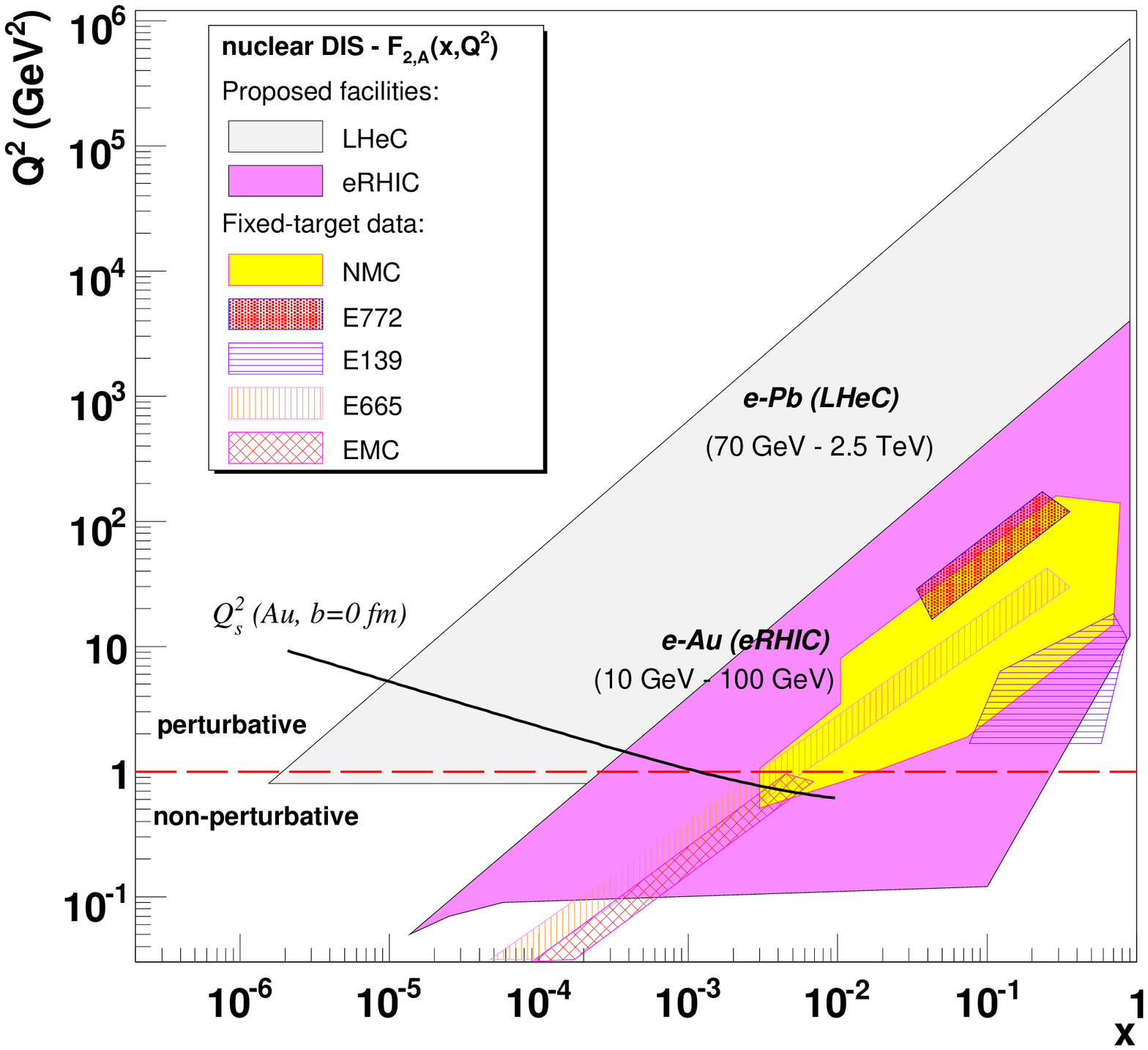,height=8cm}
\caption{Kinematic ($x,Q^2$) plane probed in e-,$\gamma$-A processes: existing data
compared to (a) UPC $\ups$ photoproduction processes (left), and
(b) new proposed nuclear DIS facilities (right): LHeC and EIC/eRHIC.}
\label{fig:nDIS}
\end{figure}

\noindent
The advance in the study of low-$x$ QCD phenomena will be specially substantial in Pb-Pb 
collisions. 
In saturation models, there is a one-to-one correspondence between the effects of rapidity- and 
$\sqrtsnn$-dependences, because a parton distribution boosted to higher rapidity $y$ is equivalent 
to a distribution sampled in a process at higher $\sqrtsnn$. As a consequence, the saturation physics 
explanation of the increasing forward suppression of semi-hard yields at RHIC
will imply a significant A-A hadron suppression at LHC {\it mid-rapidities}. 
CGC predictions for charged hadron multiplicities in central Pb-Pb at 5.5 TeV~\cite{albacete07} 
are $\dNdeta\approx$ 1500, i.e. 3-4 times lower than the pre-RHIC era results.
[As a matter of fact, such low multiplicities help CMS~\cite{cms_hi_ptdr} and ATLAS~\cite{atlas_hi} 
to become very competitive experiments in the heavy-ion running mode].
Arguably, one of the cleanest way to study the low-$x$ structure of the Pb nucleus at the LHC 
is via Ultraperipheral collisions (UPCs)~\cite{upc_yr} in which the strong electromagnetic fields 
(equivalent flux of quasi-real photons) generated by the colliding nuclei can be used for photoproduction 
studies at maximum energies $\sqrtsgN\approx$ 1 TeV, 3-4 times larger than at HERA. Full 
simulation+reconstruction studies~\cite{cms_hi_ptdr,dde_qm06} of quarkonia photoproduction
($\gpb\rightarrow\ups \,$Pb) tagged with very-forward neutrons, show that CMS can carry out 
detailed $p_T$,$\eta$ measurements in the dielectron and dimuon $\ups$ decay channels. 
Such processes probe $x$ values in the nucleus as low as $x\sim 10^{-4}$ (Fig.~\ref{fig:nDIS}, left). 

\section{Proposed future deep-inelastic facilities}

Two different collider projects -- the Large Hadron Electron Collider (LHeC)~\cite{LHeC} and the 
Electron Ion Collider (EIC)~\cite{eRHIC} -- have been recently proposed to study
deep-inelastic lepton-hadron (e-p, e-d and e-A) scattering for momentum transfers 
$Q^2$ as large as 10$^6$ GeV$^2$ and for Bjorken $x$ down to the 10$^{-6}$. Both projects
have a strong focus on the study of the low-$x$ gluon structure of protons and nuclei and on non-linear QCD evolution.
LHeC (EIC/eRHIC) proposes to add an extra $E_e$ = 70-GeV (10-GeV) electron ring to the LHC (RHIC/JLab)
proton/nucleus collider(s). In the nuclear DIS sector, LHeC (EIC/eRHIC) would allow e-A collisions 
at $\sqrtseN = 2\sqrt{E_e\,E_{_{N}}}$~=~880~(63)~GeV. Fig.~\ref{fig:nDIS} right, shows the 
($x,Q^2$) ranges accessible to both machines. 
Both proposed facilities would significantly extend the (meager) kinematical regime of the
existing nuclear DIS data, fully mapping out the range of Bjorken-$x$ at virtualities around the 
saturation scale ($Q_s^2\approx$ 1 - 10 GeV$^2$, indicated by the black curve in the plot) 
and providing very valuable insights on the high-energy limit of QCD.


\section*{Acknowledgments}

It is a pleasure to thank the organizers of Moriond'07 for their kind invitation to this unique meeting. 
Special thanks due to Fran\c{c}ois Gelis, Emmanuelle Perez, Alexander Savin, Raju Venugopalan and Graeme Watt 
for informative discussions and useful suggestions. 
This work was supported by the 6th EU Framework Programme contract MEIF-CT-2005-025073.


\end{document}